\documentclass[preprint,aps,amsmath,showkeys,showpacs,floatfix]{revtex4}
\usepackage{graphicx}% Include figure files
\usepackage{dcolumn}% Align table columns on decimal point
\usepackage{bm}% bold math

\begin{document}
%
%    figures   bb041N.eps     N=1,2,...
%
%
% *********************************************************
%   linea anyadida para que reduzca espacio entre lineas
% *******************************************************
%\tightenlines
% ********************************************************

\title{Ordered structures in rotating ultracold Bose gases}
\author{N. Barber\'an$^1$, M. Lewenstein$^{2,3}$, K. Osterloh$^3$, and D.
Dagnino$^1$}
\affiliation{$^1$Dept. ECM,
Facultat de F\'\i sica, U. de Barcelona,
E-08028 Barcelona, Spain}
\affiliation{$^2$ICREA and ICFO--Institut de Ci\`encies Fot\`oniques,
Av. del
Canal Ol{\'\i}mpico s/n, 08860 Castelldefells, Barcelona, Spain}
\affiliation{$^3$Institute for Theoretical Physics, University of
Hannover, Appelstrasse 2, 30167 Hannover, Germany}

\vskip11mm

\medskip

\begin{abstract}

The
characterization of small samples of cold bosonic atoms in rotating
microtraps has recently attracted
increasing interest due to the possibility to deal with a
few number of particles per site in optical lattices.
In this paper we consider  two-dimensional systems of few cold Bose
 atoms confined in a
harmonic  trap in the $XY$ plane, and submitted to strong rotation
around the $Z$ axis. By means of exact diagonalization,
we analyze the evolution of ground state structures
as the  rotational frequency $\Omega$ increases. Various kinds
of  ordered structures are observed. In some cases, hidden
interference patterns exhibit themselves  only in the pair
correlation function;  in some other cases  explicit
broken-symmetry structures appear that modulate the density.
For $N<10$ atoms, the standard scenario, valid for large
sytems (i.e.  nucleation of vortices into an Abrikosov
lattice, melting of the lattice, and subsequent appearance of
fractional quantum Hall type states up to the Laughlin state), is
absent for small systems, and only gradually recovered as $N$
increases.
On the one hand, the Laughlin state in the strong rotational regime
contains  ordered structures much more similar to a Wigner crystal,
or a molecule than to a fermionic quantum liquid. This result has
some similarities to electronic systems, extensively analyzed
previously. On the other hand, in the weak rotational regime, the
possibility to obtain equilibrium states, whose density reveals
an array of vortices, is restricted to some critical values
of the rotational frequency $\Omega$.
The vortex contribution to the total angular momentum $L$ as a
function of $\Omega$ ceases to be an increasing function of
$\Omega$, as observed in experiments of Chevy {\it et al.}
\cite{che}.
Instead, for small $N$, it exhibits a sequence of peaks showing
wide minima at the values of $\Omega$, where no vortices appear.

\end{abstract}
\vskip2mm

\maketitle
\vfill
\eject

\vskip11mm

\medskip

\section{Introduction}

\subsection{Ordered structures in ultracold gases and their
detection}

Ordered structures, and in particular hidden ordered structures
have been a subject of
intensive studies in the physics of Bose-Einstein condensates
(BEC's) \cite{and,kett,bra,pit}, and more generally,
in the physics of ultracold atoms.

\medskip

The paradigm example of such structures is realized in the
interference of two BEC's, observed in seminal
experiments of Ref.~\cite{and1}. Suppose that, despite the
superselection rule, one could prepare the two
condensates in coherent atomic states, characterized by fluctuating
atom numbers
$N_1$, $N_2$, but sharply defined phases $\phi_1$, $\phi_2$,
minimizing the Heisenberg uncertainty relation for the
number and phase operators. Then, the phase difference
$\Delta\phi=\phi_1-\phi_2$ would determine
the position of the interference fringes. Similarly, if we prepared
two condensates by, say, splitting
a parent condensate  with fixed number of atoms $N=N_1 + N_2$,
we would arrive at sharp values of both $\Delta\phi$ and
$N_1 - N_2$. Amazingly, the interference pattern will also appear
if the two BEC's  in the Fock states
(with fixed $N_1$ and $N_2$) overlap. The reason is, as pointed
out in Ref.~\cite{jav,dali}, that as soon as
we start detecting atoms without knowing which condensate
they originate from, the measurement will introduce
the necessary uncertainty of the atom numbers, narrowing the
relative phase distribution. As a consequence, an interference
pattern with a sharply defined $\Delta\phi$
is obtained in each realization  of the measurement.
We may say that the measurement process uncovers the otherwise
 hidden interference pattern in the two-point first
order correlation function of atomic creation and annihilation
 field operators,
$\langle \hat\Psi^\dag({\bf r})\hat\Psi({\bf r'})\rangle$.
If experimentally averaged over many realizations
the interference pattern vanishes,
since each realization leads to a different and completely random
 $\Delta\phi$. Similar measurement induced structures, and the
interplay between single shot and averaged results have also been
discussed in the context of dark solitons
in BEC \cite{sacha}.

\medskip

Other types of ordered structures occur in rotating BEC's.
In the standard scenario, as the rotational frequency increases,
more and more vortices appear in form of regular structures
\cite{pit,mad,abo}. As their number grows, they organize themselves in a
triangular Abrikosov lattice \cite{abr}. Note, that in principle
the ground state of the rotating system in a harmonic trap
should ideally be rotationally invariant and have a fixed total
angular momentum $L$, {\it ergo} it should not exhibit
any structures that break rotational symmetry, as the Abrikosov
lattice does.
In reality though, the preparation of vortices
is performed by a ``laser stirring'' process that breaks rotational
symmetry, and introduces significant
couplings between states with different total angular
momenta \cite{footnoteone}.
Here, one deals with a situation in which the preparation
process (which may also be regarded as a form of measurement)
reveals elsewise hidden structures in the density of the condensate,
i.e., in the one-point first order correlation function
$\langle \hat\Psi^\dag({\bf r})\hat\Psi({\bf r})\rangle$.

\medskip

As it is very well known from quantum optics, measurements of
first order correlation functions
(first order ``coherence'') do not always reveal the underlying
structures. In order to see them, one has to measure
higher order coherences, such as second order correlation functions
$\langle \hat\Psi^\dag({\bf r}_1) \hat\Psi^\dag({\bf r}_2)\hat
\Psi({\bf r}_3)\hat\Psi({\bf r}_4)\rangle$.
The paradigm example for this necessity goes back to Michelson
interferometry \cite{miche} which measures first
order coherences and is sensitive to atmospheric fluctuations.
This deficiency of Michelson interferometry has stimulated
Hanbury Brown and Twiss \cite{hanb}
to measure the intensity-intensity correlations of the radiation
coming from Sirius, which in turn allowed them to
precisely determine the coherence length and the angular size of this
star.

\medskip

Measurements of second order correlations play an important
role in the physics of ultracold gases (for
earlier works on atomic beams, see \cite{shim}).
The most directly measurable quantity is the density-density
correlation (pair correlation function, called pc function below):
$\langle\hat\Psi^\dag({\bf r}_1) \hat\Psi^\dag({\bf r}_2)\hat\Psi
({\bf r}_2)\hat\Psi({\bf r}_1)\rangle$,
which formally is the two-point second order correlation function
of the atomic field operators.  This function
has been directly measured in a recent atom counting experiment
of the Orsay group \cite{allain},
for the first time directly demonstrating an atomic Hanbury
Brown-Twiss effect for thermal atoms and the second order coherence
of a BEC.
Earlier, a 4-point second order correlation function
has been measured in Hannover \cite{arlt}
where density-density correlations of {\it
interfering condensates} have been monitored
in order to precisely determine the phase coherence length
of quasi-1D condensates, in full analogy to the Hanbury Brown
and Twiss method.

\medskip

Recently, yet another tool, i.e., {\it noise interferometry} has been
proposed to analyze visible and hidden structures  appearing in
various quantum phases of ultracold gases \cite{pol,bach}.
This method also allows to determine density-density
correlations, and has been used by several groups to study, for
instance,  interference of independent BEC's \cite{had}, residual
coherence and lattice order in Mott insulators \cite{foe}, and pair
correlations of fermionic atoms in a Fermi superfluid \cite{Debbie}.

\medskip

At this point it is necessary to mention that the double (spatial
and temporal) Fourier transform of the pc function is known to be a
dynamical structure factor \cite{pit}, and is also measurable, for
instance in Bragg scattering experiments \cite{sta}.

\subsection{Rapidly rotating ultracold gases}

Recently, a considerable interest has been devoted to rapidly
rotating ultracold gases, which also exhibit various kinds of
ordered structures, and should therefore be investigated along the
lines discussed in the previous subsection.

\medskip

Typically, one considers a quasi 2D gas in the $XY$ plane rotating
around the $Z$ axis with frequency $\Omega$, and confined in a
harmonic trap of frequency $\omega_{\perp}$. As stated above,
in macroscopic atomic clouds for moderate $\Omega <\omega_{\perp}$,
the Abrikosov vortex lattice is formed \cite{mad,abo}. As $\Omega$
approaches $\omega_{\perp}$, the vortex lattice melts,
and the system evolves through a sequence of strongly correlated states
\cite{wil,belen}. Finally, in the regime of critical rotation, it forms a
bosonic Laughlin liquid \cite{lau}.

\medskip

Alternatively, the various regimes of rapidly
rotating gases can be described in the terminology of fractional
quantum Hall effect (FQHE) theory \cite{book-fqhe}.
The crucial role is played by the direct analog of the Landau
level filling factor in the FQHE which can be related to the number
of vortices $N_v$ by $\nu=N/N_v$ as defined in the BEC mean field
description valid for large systems and moderate rotation.

\medskip

The first papers on atomic systems \cite{wil,belen} have
considered
the lowest Landau level (LLL) for strong enough rotation. Recently,
correlated liquids at $\nu=k/2$ for $k=1,2,3,\ldots$ for
$\nu\le \nu_c\simeq 6-10$ have been discussed \cite{coo}.
These states resemble to a great extent the states  from the
Rezayi-Read (RR) hierarchy \cite{rr}: $k=1$ is the Laughlin state,
$k=2$ is the Moore-Read paired state \cite{moore} etc.
It has been shown that the presence of but a small amount of
dipole-dipole interactions unambiguously makes the RR state with $k=3$
the ground state at filling $\nu=3/2$.
This state is particularly interesting, since its excitations are
both fractional and non-Abelian.
The validity of the LLL approximation for rotating gases is
also discussed in the recent preprint \cite{last}.

\medskip

Most of the literature on ultracold rapidly rotating gases aims at
considering relatively large systems and even the thermodynamic
limit. In numerical simulations, either periodic (torus) or
spherical boundary conditions are used.
Unfortunately, in the $N\to\infty$ limit the gap separating the
Laughlin state from its excitations vanishes. Observation of
Laughlin states not only requires
to reach the LLL, but also to control very precisely a delicate
 balance between $\Omega$ and $\omega_{\perp}$.
Despite the progress in experimental studies of
vortex lattices \cite{cornell,jean1}, and first steps towards
LLL physics \cite{jean2}, experiments have
not yet reached this regime.

\medskip

The problems related to the short range nature of the Van der Waals
forces can be overcome in dipolar gases, i.e. gases that interact
via magnetic or electric dipole moments
(for a review see \cite{baran}). Rotating dipolar bosonic gases are
expected to exhibit exotic behaviour
in the weakly interacting regime \cite{cooper-weak}, whereas
fermionic dipolar gases  have a finite gap for
the $\nu=1/3$ Laughlin state \cite{oster}. The first observation
of BEC of a dipolar gas of  Chromium atoms
with large magnetic dipole has been recently reported \cite{pfau},
and several groups are trying to realize and control an ultracold
gas of heteronuclear molecules with large electric dipole moments
\cite{molec}.

\medskip

Another way to create highly-correlated liquids could be,
not to mimic effects of magnetic fields by rotation,
but by appropriately designed control of tunnelling phases
in optical lattices \cite{jaksch}. In trapped gases, a similar
effect may be realized by employing electromagnetically induced
transparency \cite{fleisch}.

\medskip

However, the most promising way towards the FQH regime and related
states may be achieved by use of an array of rotating optical
microtraps, either in an optical lattice \cite{bloch-pri}, or
created by an array of rotating microlenses
 \cite{birkl}.

\medskip

In such arrangements, it will be natural to study mesoscopic, or even
microscopic systems of few atoms.
Such experiments demand careful theoretical studies of few atom
systems using possibly exact methods, such as exact
diagonalizations of the Hamiltonian with open boundary conditions
in  the presence of the harmonic trap, or even a deformed trap.
Such studies have recently been initiated \cite{cirac}, and the
possibilities of an adiabatic path to fractional quantum Hall states
of a few bosonic atoms have been investigated in detail. We
continue the studies of small systems of atoms in rotating traps,
and expand them in the present paper.

\subsection{Plan of the paper}

The main focus of the present paper is to study and analyze the
ground state (GS) ordered structures and interference patterns
(IP) of two-dimensional Bose systems of few atoms confined in a
harmonic trap and submitted to fast rotation around the
perpendicular axis. We investigate here, on one hand, the situations
where the cylindrical symmetry is explicitly broken
so that the one particle density already exhibits
ordered structures due to the coherent mixing of degenerated GS
with different total angular momenta. Furthermore,
we consider situations in  which the ordered
patterns are hidden in a pure single state with well
defined angular momentum,
and are evident only through inspection of the  pair correlation
(pc) function calculated by means of exact diagonalization. This
formalism turns out to be the appropriate method to deal with small
systems, for which the assumptions made in mean field theories do
not apply. Such systems are experimentally accessible, and both,
density and pc functions, are measurable by various experimental
techniques discussed in the previous subsection.
Experimental information on the IP can be obtained in the last
case.

\medskip

According to our findings, the behaviour of confined systems of few
atoms strongly differs from the behaviour of large systems. These
differences are not only related to the nucleation of
vortices in the regime of relatively slow rotation, but also
to the nature of the Laughlin state and other
highly-correlated states, when the rotational frequency is close
to the trap frequency.

\medskip

In particular, we obtain that for $N<10$ atoms the
standard scenario valid for  large sytems (i.e., the nucleation of
vortices into an Abrikosov lattice, melting of the lattice,
and subsequent appearance of fractional quantum Hall type states up
to the Lauhglin state),  is absent for small systems, and is only
gradually recovered as $N$ increases.
On the one hand, the Laughlin state in the strong
rotational regime  contains  ordered
structures much more similar to
a Wigner crystal or a molecule than to a
Fermi liquid. This result has  some similarities to electronic
systems, extensively analyzed previously. On the other hand,
in the weak rotational regime, the possibility to obtain
equilibrium states whose density reveals an array of
vortices is restricted to some critical values of the rotation
frequency $\Omega$.
The vortex contribution to the total angular momentum $L$ as a
function of
$\Omega$
ceases to be an increasing function of $\Omega$, as observed in
experiments of Chevy {\it et al.}
\cite{che}. Instead, for small $N$, it exhibits a sequence of
peaks showing wide minima at the values of $\Omega$,
where no vortices appear.

\medskip

This paper is organized as follows. In Sec.~II we describe our
system (Section IIA) and address the questions
related to the realization and analysis of ground states with
hidden (Section IIB), or explicit (Section IIC)
broken cylindrical symmetry. In Sec.~III,
the main results of this work are presented. Finally, in
Sec.~IV, we compare our findings with previous results in the
literature, and draw our conclusions.

\section{Ordered structures in ground states: broken cylindrical
 symmetry}

\subsection{Description of the system}

Our system consists of $N$ bosonic atoms trapped in a rotating
parabolic potential. The Hamiltonian in the rotating reference
frame reads \cite{caz},

\begin{equation}
H= \sum_{i=1}^N \left [ \frac{(\vec{p}-\frac{e}{c}\vec{A^*})_i^2}{
2M}
+
\frac{1}{2}M(\omega_{\perp}^2-\Omega^2) r_i^2\right ] + g\sum_{i<j}
\delta(\vec{r}_i- \vec{r}_j),
\label{Hamiltonian}
\end{equation}
where $\vec{r}=(x,y)$, $\omega_{\perp}$ is the trap frequency,
$\vec{A^*}
=
\frac{M\Omega c}{e}
\hat{z}\times\vec{r}\,\,\,$ is the vector potential,
$\,\,\,\hat{z}\,\,$ is the unitary vector along the $Z$ direction
and $\,\,\vec{B^*}=\vec{\nabla}\times\vec{A^*}\,\,=\,\,\frac{
2M\Omega c}{e}\hat{z}\,\,\,$ is the effective magnetic field of an
equivalent system of electrons submitted to a magnetic
field perpendicular to the $XY$ plane (we use here the symmetric
gauge).
Thus, the rotation
of the trap has formally the same effect on atoms of mass $M$, as
a magnetic field has on electrons; the electronic
charge $-e$ and the speed of light $c$ are solely introduced
for reasons of algebraic equivalence.
$V=g\sum_{i<j}\delta (\vec{r}_i-\vec{r}_j)$ is the contact
interaction potential, where $g$ is
the interaction coefficient that approximates  the potential of the
Van der Waals forces between the atoms in the very dilute limit.
We assume the rotational frequency to be large enough to restrict
the system to the lowest Landau
level (LLL) regime, and choose the appropriate Fock--Darwin single
particle (sp) wave functions with no nodes in the radial
direction, as the basis in order to represent all operators
\cite{jac},
\begin{equation}
\mid m\rangle =\frac{1}{\lambda\sqrt{\pi
m!}}\,\,\left(\frac{z}{\lambda}\right)^m
\,\,e^{-\mid z\mid^2/ 2 \lambda^2}
\label{5}
\end{equation}
with $\lambda=\sqrt{\frac{\hbar}{2M\omega_{\perp}}}$,
and generalized complex coordinates $z=x+iy$.

\medskip

The Hamiltonian can be written in second quantized form as,
\begin{equation}
\hat{H}= \alpha \hat{L} + \beta \hat{N} + \hat{V},
\label{secondquant}
\end{equation}
where
$\alpha=\hbar ( \omega_{\perp}-\Omega)$,
$\beta=\hbar\omega_{\perp}$,
$\hat{L}$ and $\hat{N}$ are the total $z$-component angular momentum
and particle number operators, respectively, and
\begin{equation}
\hat{V}=\frac{1}{2}\sum_{m_1m_2m_3m_4}
V_{1234}\,\,\,a^\dag_1a^\dag_2a_4\,a_3\,\,\,\,,
\end{equation}
where the
matrix elements of the interaction term are given by
\begin{equation}
V_{1234}=\langle m_1\,m_2 \mid V \mid m_3\,m_4 \rangle=
\frac{g}{\lambda^2 \pi}\,\,\frac{\delta_{m_1+m_2,
m_3+m_4}}{\sqrt{m_1!m_2!m_3!m_4!}}\,\,
\frac{(m_1+m_2)!}{2^{m_1+m_2+1}}\: .
\end{equation}
Here, the operators
$a^\dag_i$ and $a_i$ create and annihilate a boson
with single-particle (sp) angular momentum $m_i$, respectively.
The cylindrical symmetry of the Hamiltonian allows
the diagonalization to be performed in different subspaces of well
defined total $z$-component of angular momentum $L=\sum_{i=1}^Nm_i$.

\medskip

%c
%c            110 % al hacer el *.eps
%c     Mac1.for   g=\omega_{\perp}=1    \Omega = 0.94   N=5
%%%%%%%%%%%%%%%%%%%%%%%%%%%%%%%%%%%%%%%%%%%%%%%%%%%%%%%%%%%%%%%

\begin{figure}[htb]
\includegraphics*[width=0.7\columnwidth]{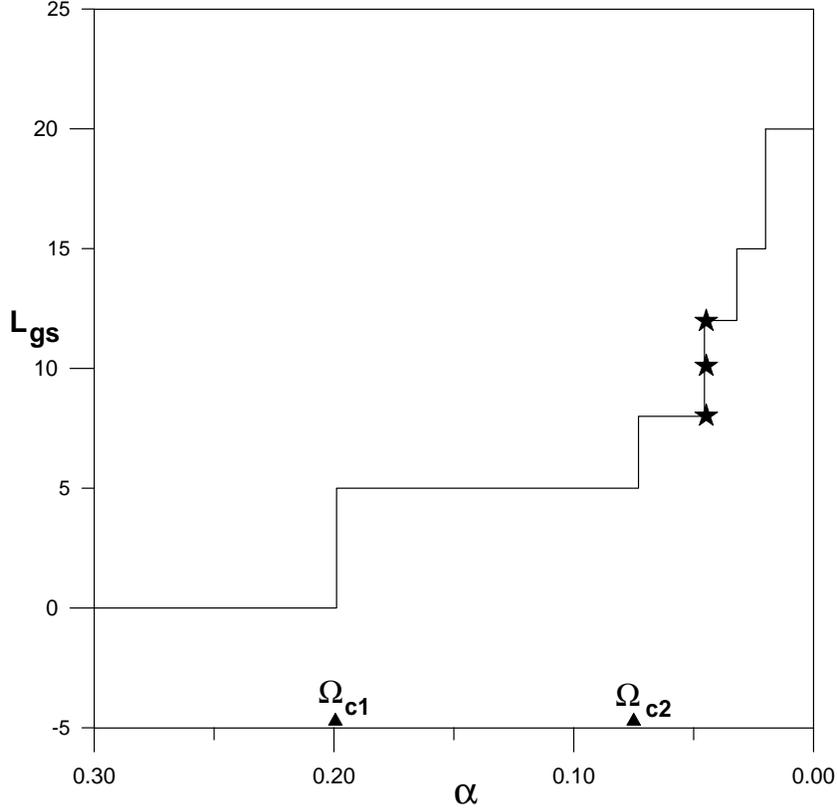}
\caption{Change of the GS angular momentum $L_{gs}$ for $N=5$ as
the rotation frequency
increases; transitions take place at critical values of the
rotational frequency labelled by $\Omega_{cn}$.
$(\alpha=\hbar(\omega_{\perp}-\Omega))$.}
\end{figure}

%%%%%%%%%%%%%%%%%%%%%%%%%%%%%%%%%%%%%%%%%%%%%%%%%%%%%%%%%%%%%%%%

Fig.~1 shows the total angular momentum of the GS of a system
of $N=5$ particles while $\Omega$ grows from zero to $\omega_{\perp}$,
the maximum possible value before the system becomes centrifugally
unstable. We observe that the GS angular momentum
remains constant for a finite range of $\Omega$ until transitions
to new angular momenta take place at critical values labelled as
$\Omega_{cn}$. Not all $L$-values can be associated with the
GS's. However, on the
steps, different L-states may be degenerate in energy as
in the case of the states with $L=8,10$ and $12$ on the third step
(indicated by stars in Fig.~1). The last
possible GS at $L=N(N-1)$ is the Laughlin state, for which the
interaction energy is zero due to the fact that the wave function
of each atom has zeros of order two at the positions of the other
$N-1$ atoms; this can be easily deduced from
the analytical expression of the many-body wave function given by,

\begin{equation}
\Psi_{Laughlin}={\cal{N}}\prod_{i<j}(z_i-z_j)^2e^{-\sum \mid z_i\mid^2/ 2
\lambda^2}\,\,\,.
\label{laughwave}
\end{equation}

\medskip
In the following, the GS of the system in the $L$-subspace is
denoted by $\Psi_L$.
In order to study the nature of the GS's, it is useful to analyze
the expectation values of some relevant operators. First of all, it
is crucial to realize that the density operator defined in first
quantization as
\begin{equation}
\hat{\rho}(\vec{r})=\sum_{i=1}^N\delta(\vec{r}-\vec{r}_i)
\end{equation}
does not exhibit any interference pattern when calculated for a
definite $\Psi_L$, as it can be inferred from its analytical
expression in second quantized form
\begin{equation}
\hat{\rho}(\vec{r})=\sum_{ij}\langle\phi_i(\vec{r'})\mid\delta
(\vec{r} -\vec{r'})\mid \phi_j(\vec{r'})\rangle a_i^\dag a_j,
\end{equation}
where $|\phi_i(\vec{r})\rangle = |m_i\rangle$ as in Eq.~(\ref{5}).
Due to angular
momentum conservation, the operator $a_i^\dag a_j$ selects only one
sp state and, as a consequence, it loses all information
contained in products of different
amplitudes, thus losing the interference pattern. It solely preserves
the information of individual densities, e.g.,
\begin{equation}
\rho(\vec{r})=\langle \Psi_L\mid\hat{\rho}(\vec{r})\mid\Psi_L
\rangle= \sum_i^N\mid\phi_i(\vec{r})\mid ^2 Oc_i \:,
\end{equation}
where $Oc_i$ is the total occupation of the sp state
$|m_i\rangle$ in the GS. In effect, $\rho(\vec{r})$  can only
represent  cylindrically symmetric
distributions. However, we note that this cylindrical symmetry is a
direct consequence of the definition  of the
operator $\hat\rho(\vec{r})$,  and is not necessarily a
manifestation of the symmetric nature of the GS.

\medskip

To exhibit ordered patterns and analyze the GS structures, we
proceed in two different ways;
one investigates the pair
correlation function for states with fixed $L$ (Section
IIB), the other combines different $\Psi_L$'s
(Section IIC).

\medskip

\subsection{Ordered structures in pair correlation functions}

In order to analyze the internal structure of relevant states,
we consider the following operator
\begin{equation}
\hat{\rho}(\vec{r},\vec{r}_0)= \sum_{i<j}^N \delta
(\vec{r}_i-\vec{r}_0)
\delta(\vec{r}_j-\vec{r}),
\label{10}
\end{equation}
which yields the
conditional probability to find an atom at $\vec{r}$, when
another is simultaneously found at $\vec{r}_0$.
This operator contains information that originates from the
amplitudes of sp wave functions, and
not only from
their density as it has been in case of the single particle
density operator.
In second quatized formalism, its expected value
with respect to $\Psi_L$ reads
\begin{equation}
\rho(\vec{r},\vec{r}_0)=\sum_{ijkl}\sum_{pp'} \alpha_p^* \alpha_{p'}
\phi^*_i(\vec{r}) \phi^*_j(\vec{r}_0)\phi_k(\vec{r}) \phi_l(\vec{r}_0)
\langle\Phi_p\mid a_i^\dag a_j^\dag a_la_k\mid \Phi_{p'}\rangle,
\label{two-pc}
\end{equation}
where
\begin{equation}
\Psi_L=\sum_{p=1}^{n_d} \alpha_p \Phi_p,
\end{equation}
and $\Phi_p$ are the bosonic Fock $N$-body states of the basis
in the $L$-subspace of dimension $n_d$. The condition $i+j=k+l$
must be fulfilled for reasons of angular momentum conservation.
It should be stressed that $\rho(\vec{r},\vec{r}_0)$
in Eq.~(\ref{10}) obviously differs from the single particle
density matrix
\begin{equation}
n^{(1)}(\vec{r},\vec{r'})=\langle\hat{\Psi}^+ (\vec{r}) \hat{\Psi}
(\vec{r'})\rangle,
\end{equation}
which defines the off-diagonal long-range order
that characterizes Bose condensation \cite{pit}. The
operator
$\hat{\rho}$ is a two-particle operator, whereas
$\hat{n}^{(1)} (\vec{r},\vec{r'})=\hat{\Psi}^+(\vec{r})
\hat{\Psi}(\vec{r'})$ is a single-particle operator; in  particular,
$\rho(\vec{r})=n^{(1)}(
\vec{r},\vec{r})$, whereas $\rho(\vec{r})=\frac{1}{N-1}\int
d\vec{r}_0
\rho(\vec{r},\vec{r}_0)$. As a rule of thumb, if
$\hat{n}^{(1)} (\vec{r},\vec{r'})$ reveals symmetry breaking, so
does $\hat{\rho}$, whereas the opposite is not necessarily true.
\medskip

\medskip

Eq.~(\ref{two-pc}) can be interpreted as the sum of products of
amplitudes at $\vec{r}$
weighted by a factor that depends on $\vec{r}_0$, and
on the GS via the $\alpha_p$ coefficients. In the particular case
$\vec{r}_0=\vec{0}$, cylindrical symmetry is recovered, since in
this case $l=j=0$ is the unique non-zero contribution. This implies
$i=k$ and yields
\begin{equation}
\rho(\vec{r},\vec{0})=\mid\phi_0(\vec{0})\mid^2
\sum_i\mid\phi_i(\vec{r})\mid^2
\sum_{pp'}\alpha_p^* \alpha_{p'} \langle ... \rangle,
\end{equation}
which is independent from $\theta$.

\medskip

In order to understand the role of the parameter $\vec{r}_0$ in
 $\langle \Psi_L\mid \hat{\rho}(\vec{r},\vec{r}_0) \mid
\Psi_L\rangle$ as a function of $\vec{r}$, we consider
\begin{equation}
\phi^*_n(\vec{r})\phi^*_j(\vec{r}_0)\phi_k(\vec{r})
\phi_l(\vec{r}_0) =
\frac{1}{\pi^2}\frac{r^{m_n}}{\sqrt{m_n!}}
\frac{r^{m_j}_0}{\sqrt{m_j!}}
\frac{r^{m_k}}{\sqrt{m_k!}}
\frac{r^{m_l}_0}{\sqrt{m_l!}} e^{i(m_k-m_n)\theta}
e^{i(m_l-m_j)\theta
_0} e^{-r^2} e^{-r_0^2},
\end{equation}
in units of $\lambda$. As  $l-j=n-k$ follows from angular momentum
conservation, the angular dependence reads
\begin{equation}
e^{i(m_k-m_n)\theta} e^{i(m_l-m_j)\theta_0} =
e^{i(m_k-m_n)(\theta-\theta_0)}.
\end{equation}
Evidently, if $r_0$ is fixed, the change of $\theta_0$ is
nothing but a rigid rotation of the function. In other
words, any arbitrary choice of $\theta_0$ fixes the origin of
angles, and breaks cylindrical symmetry,  in the
analogous way as it happens in experiments which perform
a single shot measurement. Within this point of view, the experimental
measurement and the choice of $\theta_0$
are equivalent processes (see for instance \cite{foe,pol}).

\medskip

The expected values of the pc function for
$r_0\neq 0$ can reveal very different situations:  from circular
symmetric structures showing no spatial correlation,  to ordered
structures that reveal intrinsic Wigner molecules or crystals
while passing through all possible intermediate states, as
it is shown in Section III.

\medskip

\subsection{Ordered structures in the density: superpositions  of
different L-subspaces}

In this subsection, we consider ordered structures in GS's with no
well defined angular momentum in two different situations.
On the one hand, we build (somewhat {\it ad hoc}) linear
combinations of different $\Psi_L$'s to explicitly reveal the structure
present in the expectation value of the density operator. We
obtain ordered IP's for combinations, whenever one of the
contributing $\Psi_L$'s has an ordered hidden IP contained in its
pc function.
On the other hand,
after the introduction of an anisotropic term to the Hamiltonian which
mimics the deformation introduced by the stirring laser, we
perform numerical diagonalization without the restriction of angular
momentum conservation, and  obtain in this way GS structures with
broken symmetry. These exact calculations give hints, how to
construct approximated GS superpositions in the previous
{\it ad hoc} construction.

\medskip

To explain more precisely, how the first procedure works, we
start from the pc function and observe what
kind of ordered structures can be expected.
In case of the Laughlin state $L=20,N=5$, the pc
function suggests that the atoms form a pentagon.
Quite generally, the best way to visualise this structure
within the first procedure is to form a superposition
$A\Psi_L + B\Psi_{L+N}$, where $\Psi_L$ is the GS that
contains a hidden
ordered IP. It
is easy to understand this result from what follows. The terms
that contribute to the broken cylindrical symmetry
are those of the form $\langle
\Psi_L\mid a^\dag_i a_j\mid \Psi_{L+M}\rangle$ with $M\geq 1$.
However, it is necessary to arrive at $M=N$ in order to obtain
contributions from all the sp states contained in $\Psi_L$. To be
more precise, none of the combinations from $L=20+21$ to $L=20+24$
reproduces the structure of $L=20$ for $N=5$. It is necessary to
combine $L=20+25$ to obtain the regular pentagon implicit in
$\Psi_{20}$.
The best contrast is obtained for
$A=B=1$, and not for a small amount of
$\Psi_{L+N}$ as one would expect if only a perturbation would be
necessary.

\medskip

It is important to stress, that the ordered hidden IP
was obtained in
the Laughlin state $L=20$ for $N=5$.  In order
to assure that this state is the GS, a very small amount of kinetic
energy is necessary in such a way that the states $L+M$ for
$M\ge 0$ are quasi-degenerate.
Then, the combination considered above corresponds to a
``legitimate'' GS.

\medskip

The second procedure,
followed in order to obtain ordered structures as, e.g., multiple
vortex states, was suggested by
experiments. In the experimental setup described by
Chevy {\it et al.} \cite{che} and by Madison {\it et al.}
\cite{mad}, vortices are
generated as equilibrium states of a Bose condensate rotating under
the action of a stirring laser that produces anisotropy in the
$XY$ plane. Subject to this anisotropic potential, the state with
vortices is a GS, and it survives during
the time of flight (TOF) detection as an excited state of the
restored symmetric Hamiltonian after the trap is switched off.

\medskip

With this idea in mind, we introduce an additional anisotropic term in
the Hamiltonian given by $\hat{V}_p=A\sum_{i=1}^N(x^2-y^2)_i$ or in
second quantized form as \cite{footnotetwo}
\begin{equation}
\hat{V_p}=\frac{A}{2}\lambda^2\sum_m\left
[\sqrt{m(m-1)}a^+_ma_{m-2}+
\sqrt{(m+1)(m+2)}a^{\dag}_ma_{m+2}\right ]\,\,.\label{aniso}
\end{equation}
We assume this term to be a small perturbation of the
system, thus,
$\frac{A\lambda^2/2}{\hbar(\omega_{\perp}-\Omega)}<<1$,
and perform exact diagonalization to obtain the GS of $H+V_p$.

\medskip

Amazingly, the structure of, say, two vortices
can only be obtained at very specific plateau steps $\Omega_{cn}$.
There, the GS is a combination of
quasi-degenerate $\Psi_L$-states which are coupled by a perturbing term
slightly larger than their energy difference but much smaller than the
next eigenenergy. As a consequence, the linear combination of
the states above has nearly equal coefficients.
Thus, a direct {\it ad hoc} combination of degenerated states of the
symmetric Hamiltonian of Eq.~(\ref{Hamiltonian}) are educated guesses
to reveal underlying structures.
This combination was previously used by Wilkin {\it et al.}
in exact diagonalization calculations to obtain two vortices
\cite{wil}.
More precisely, the unique situation where vortices
are generated in the density
corresponds to the steps in the $L_{gs}$ dependence on $\Omega$,
where a degeneracy of states with different $L$ takes place
at $\Omega_{cn}$.
At first sight,
this result does not agree with the experimental results reported
by Chevy {\it et al.} \cite{che}. However, it
can be attributed to a essentially different behaviour of
systems with a large and a small number of atoms, respectively.
As $N$ grows, the size of some of the plateaus shown in Fig.~1
drastically
shrinks in such a way that finite ranges of $\Omega$-values with
energetically degenerate states become possible;
not only at critical values $\Omega_{cn}$.
In Fig.~2, we show the appearance of such
microplateaus obtained
for $N=6$, $7$, $8$, and $9$.

\medskip

Deduced from these features, our prediction is that the experimental
graph, analogous to the one displayed by Chevy {\it et al.}
\cite{che}
in their Fig.~2 which shows a monotonous growth of the vortex
contribution to $L_{gs}$ as a function of
$\Omega$, from the first vortex nucleation (at $\Omega_{c1}$) to the
turbulent regime, would look radically different for  small $N$. We
expect that it would present a curve with minima at those values of
$\Omega$ where the GS is deep in a plateau,  and exhibits no
vortices in the density.
%%%%%%%%%%%%%%%%%%%%%%%%%%%%%%%%%%%%%%%%%%%%%%%%%%%%%%%%%%%%%%%%%%%%
Vortices will solely be visible on
microplateaus surrounding $\Omega_{cn}$. For  fixed but larger
$N$, the microplateaus contain more and
more states as $\Omega$ increases,  and thus a larger  number of
vortices is nucleated. Ultimately, in the large $N$ limit,
the number of vortices becomes proportional to
the rotational frequency.

\medskip

%%%%%%%%%%%%%%%%%%%%%%%%%%%%%%%%%%%%%%%%%%%%%%%%%%%%%%%%%%%%%%%%%

\begin{figure}[htb]
\includegraphics*[width=0.7\columnwidth]{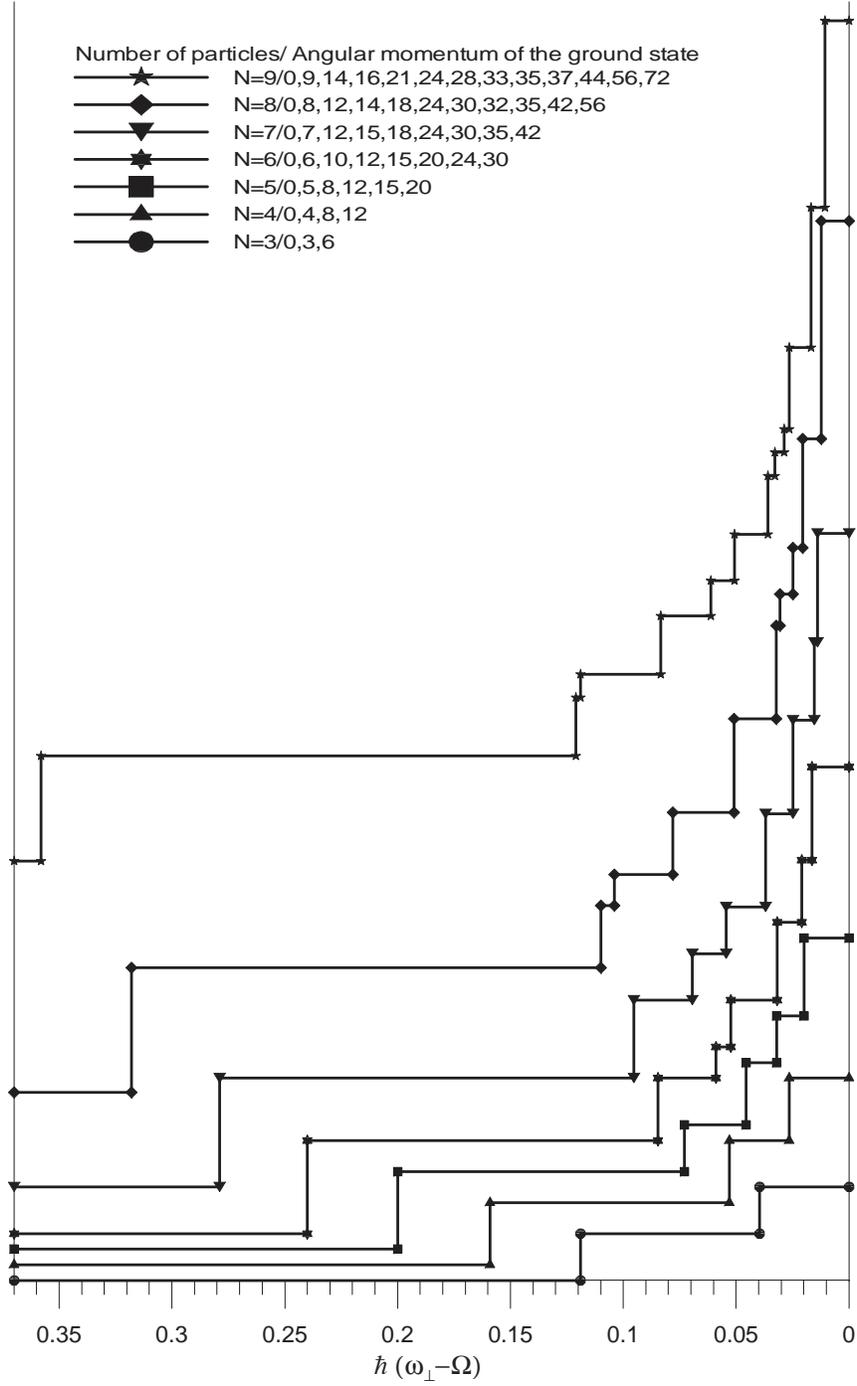}
\caption{The same as in Fig.~1 for $N=3$ to $9$ from bottom to top.
Graphs are vertically shifted for clarity. The (quasi-)degenerate
states within the same step are not included. }
\end{figure}
%\eject

%%%%%%%%%%%%%%%%%%%%%%%%%%%%%%%%%%%%%%%%%%%%%%%%%%%%%%%%%%%%%%%%

\section{Numerical results}

In what follows, we display the results obtained
from exact diagonalization for $g\lambda^2=1$ in units of
$\hbar\omega_{\perp}$. The values
$\omega_{\perp}=
\Omega=0$ were considered for density and correlation functions of
single states, or combinations of states with well defined angular
momentum, as the diagonalization of $H$ depends only on the
interaction energy. In contrast, specific values of $\alpha$ and
$\beta$ (see Eq.~(\ref{secondquant})) are considered when the
diagonalization of $H+V_p$
is performed. Correlation functions are always displayed
in pairs, a 3D and a contour plot, unless otherwise specified. We
consider $\lambda$ and $\hbar \omega_{\perp}$ as units of length and
energy unless otherwise stated.

\medskip

For $N\gg 1$, the first vortex is nucleated at
$\Omega_{c1}= \omega_{\perp}-\frac{gN}{8\pi}$ when the
transition from
$L=0$ to $L=N$ takes place at the first step of
$L_{gs}(\Omega)$. The $L=N$ state whith a vortex
at the centre is the
GS from $\Omega_{c1}$ to $\Omega_{c2}$ (see Fig.~1). This state is
characterized by a high occupancy of the $m=1$
sp state and circular symmetry possessing no space
correlations. In contrast, for small values of $N$ a vortex is not
clearly manifested unless a considerably large number of atoms is
considered, as it is shown in
Fig.~3, where the pc of
the $L=N$ state is displayed
for $N=3$, $10$ and $20$. In all the cases, the density has a
minimum at the origin and $r_0$ is set to its maximum.
The slow tendency to recover the behaviour
of large condensed systems is explained in Fig.~4, where the
occupation of the
$m=1$ sp state over $N$ is shown. Namely, for few atoms, the angular
momentum
of the state is not fully due to vortices. For a large number of
atoms, Nakajima and Ueda have analyzed the formation of the first
vortex, from its initial nucleation at the cloud boundary towards
its final stabilization at the centre \cite{nak};
our results share some similarity with this
situation: for small $N$, the vortex is not yet fully inside the
trap, and as $N$ increases, it approaches
the trap centre from the boundary of the cloud.

\medskip

In Fig.'s~5 to 9, we show the main results for $N=3$ including the
ground state evolution
as $\Omega$ increases. Fig.~5a shows the lowest eigenenergies for
each $L$, the so-called Yrast line. The initial points
of the plateaus, at
$L=0,3$ and $6$ are the unique possible GS's (besides degeneracies
at the steps).
A general result is that the
plateau previous to the Lauhglin state (from $L=3$ to $L=5$ in this
case) has always $N$ points.
Fig.~5b is similar to Fig.~1 for $N=5$.
The densities
from $L=0$ to $9$ are shown in Fig.~6. The parameter $r_0$ used in
the pc calculation was set to the maximum of the
density unless it is located at the centre.
Then, $r_0=1$ is used if not
stated elsewise. Fig.'s~7 and 8 display the pc function for
$L=0,3,4,5,6$, and $9$. The system
evolves from a completely ``condensed'' system at $L=0$ to the
Laughlin state at $L=6$ where a clear triangular structure appears.
The loss of condensation is related to the increase of space
correlations. To complete the analysis, we show in Fig.~9 the
evolution of sp occupations, and demonstrate that
``macroscopic'' occupation of a specific sp wave function
vanishes as $L$ increases.

\medskip

In order to see how the previous general tendency evolves as $N$
increases, we analyzed the $N=5$ case. In Fig.~10 we show the
densities of the GS's from $L=0$ to the Laughlin state, and in
Fig.'s~11 and 12 we display their pc. The same tendency towards space
ordering at the Laughlin state is clear. In addition, it can be
inferred from the $L=20$ case that correlations are stronger
for nearest neighbours as
a manifestation of partial long range order in finite systems.
In Fig.~13 the occupations of sp states are shown. It is remarkable
that some indications of the Laughlin state typical for large systems,
i.e., a flat density at the central part
and a hump at the edge are already manifested in such small systems,
as it can be seen in the last graphs of Fig.~13 and Fig.~10,
respectively. Moreover, the density at
the origin is very close to $1/(2\pi)$, as necessary for a
homogeneous system at filling factor $1/2$.

\medskip

In Fig.'s~14 and 15, we concentrate on the Laughlin state for
$N=3,4,5,6,7$, and $8$ atoms.
The left hand side picture for $N=3,4$ and $5$ contains
the pc function and the right hand side displays the density
of the superposition made from $L$ and $L+N$.
This superposition of
quasi-degenerated states becomes a possible realization of a GS for a
certain value of $\Omega$, as commented previously.
The value of the
parameter $r_0$ in the pc functions was obtained in a different way
as the one used previously. Taking
advantage of the fact that the
Laughlin wave function (Eq.~(\ref{laughwave})) is the exact
solution, and knowing that its pc shows a ring shape structure of
an unknown radious $r_0$, we can maximize the probalibity
distribution given by,
\begin{equation}
\mid \Psi_{Laughlin}(\vec{r_1},\vec{r_2},...\vec{r_N}) \mid^2 = e^{-T}
\end{equation}
where
\begin{equation}
T=\sum_i \frac{r_i^2}{2\lambda^2} - 2q\sum_{i<j} \ln\mid z
_i-z_j\mid
\end{equation}
(with $q=2$ for the bosonic Laughlin state),
or equivalently, minimize $T$ with respect to $r_0$.
Minimization
yields $r_0=\sqrt{N-1}$ (or $r_0=\sqrt{N}$ if one atom is at the
origin as for $N=6$, $7$ and $8$) which is always smaller than the
size of the system given by $R=\sqrt{4N-2}$. To see the evolution
of these hidden ordered structures as $N$ increases, the
``degree of correlation'' $C$ as
a function of $N$ is displayed in Fig.~16; $C$ is defined as the
height of the maximum peak in the pc function. It
decreases for increasing $N$ as it is expected in order to recover
the quantum liquid character of the Laughlin state for large systems.

\medskip

Finally, for $N=5$ and $6$, Fig.~17 shows patterns of two
incipient vortices obtained
from full diagonalization of $H+V_p$ at
$\hbar(\omega_{\perp}-\Omega)=0.0458$ and
$\hbar(\omega_{\perp}-\Omega)=0.05904$ in units of $\hbar
\omega_{\perp}$, respectively.
The possibility to obtain these
patterns for the given anisotropy strongly depends on the
possibility to obtain truly degenerate states with angular
momenta $L$ and $L\pm 2$ at one of the steps $L_{gs}(\Omega)$.
For $N=7$ we were not able to find such steps.
For $N=5$ the result of the diagonalization at the third step,
where $L=8,10$ and $12$ are involved, is that the
weights of $L=8$ and $10$ within the
expansion of the GS are much larger that the weight of $L=12$. This
leads in effect to a state with expected vortex
angular momentum lower than $10$, in agreement with the results
demonstrated in Ref.~\cite{zam} related to the fact that the
contribution of a vortex to the total angular momentum depends
on its distance from the origin, it runs from $N$ at the centre
to zero at the trap boundary.

\medskip

\section{Discussion and summary}

The  main result of this paper is the identification of
important differences between large and small systems of rapidly
rotating cold bosonic atoms.
These differences can be understood by looking at the expected
values of the density and the pc functions, on which we have
concentrated our analysis.
The characterization of small samples in rotating traps has
recently attracted increasing interest due to the possibility
to deal with a few number of atoms per well
in optical lattices \cite{cirac}.

\medskip

Within the regime of low rotational frequency, we obtain
that a relatively large number of atoms is necessary to
nucleate the first vortex carrying $N$ units of angular
momentum. The evolution
towards the condensed state with $L=N$ is shown by the increase
of the occupation of
the sp $m=1$ state as $N$ increases. On the other hand, in the
regime of strong rotation, space correlations increase
significantly, and in the Laughlin state a hidden ordered
structure modulates the pc pattern.
For small systems, the atoms sit around a ring of radius
$r_0=\sqrt{N-1}$ ($N=3$, $4$ and $5$) or $r_0=\sqrt{N}$ ($N=6$, $7$
and $8$). The degree of correlation defined as the height
of the peaks in the ordered pattern decreases with $N$, evolving
towards a non-correlated structure of a quantum liquid. We have
argued about the observability of the ordered IP in similar
experiments as those reported, for instance,  by F\"{o}lling {\it et
al.}
\cite{foe}.

\medskip

Numerous references have analyzed the Wigner
structures of few electrons \cite{jau,yan,shi1,ras,yan1,yan2,sza},
mostly for filling factors less than $1/3$,
and exhibited in the density (and not in the pc function).
It is important to
remark that it is well established from exact diagonalization
studies in a torus geometry and from the analysis of the Laughlin
wave function, that states of filling factor $1/3$ for electrons
and $1/2$ for bosons are fermionic quantum liquids in the
thermodynamic limit (cf.~\cite{yos}).
Whenever the Laughlin function is a good approximation, its implied
properties are independent from the interaction. However, for small
confined systems the previous results do not apply, and the
analysis of some of their properties relies on the
competition between the kinetic and interaction energies, aside
from their statistics.

\medskip

Suggested by our results, a possible explanation for the realization
of Wigner molecules in the Laughlin states for few bosons is the
following. The first observation is that
the nature of the GS does not depend on the kinetic part as
the diagonalization is fully determined by the interaction, in
other words, the structure does not result from the competition
between different kinds of energy. In addition, as the repulsive
interaction energy is zero in the Laughlin state, it seems that the
reason why the atoms choose symmetric and well separated positions
is due to two conditions, firstly, the system must have a large angular
momentum given by $L=N(N-1)$ (which means large distances from the
origin) and secondly, each atom is surrounded by a quasihole
(which leads to effective mutual repulsion). This last statement is
supported by the following observation. The contour plots of $N=3$ for
$L=4$ and
$L=5$ in
Fig.'s~7 and 8 suggest that in those precursory states (the Laughlin
state has $L=6$), quasi-holes not attached to atoms are created
without cost of internal energy, the contribution to the angular
momentum of
each one would evolve as $1/3$ (in $L=4$), $2/3$ (in $L=5$) and $3/3$
until the Laughlin structure becomes possible
with one quasi-hole attached to each atom, lowering the
interaction energy.
A final observation relates to the evolution of this behaviour as
$N$ increases. Due to the fact that the dependence of
$L$ on $N$ is quadratic, the increase of $L$
with $N$ is more efficient for large $N$, and atoms
do not have to be widely separated. Thus, the symmetric distribution
tends to disappear.
Recently, it has been proposed that the phenomenology of
strongly correlated bosonic and fermionic rotating systems
converges to te case of classical particles, and finally
cristallizes at high rotational frequencies \cite{cri}.
Our results exhibit some traces of such cristallization,
but it should be pointed out that this ''cristallization''
ceases to be manifested when the bulk structure
starts to dominate the system. This happens for bigger particle
numbers, where the GS at Laughlin angular momentum will
start to behave more and more like a true quantum liquid
as pointed out above.

\medskip

Finally, precursors of two-vortex arrays are obtained as the ground
states of an asymmetric Hamiltonian that models the experimental setup
used to increase the angular momentum of a trapped Bose condensate
by a stirring laser. We conclude that the possibility to nucleate
vortex patters in the density is restricted for small $N$ to the
specific values $\Omega_{cn}$ at the steps
where several degenerate or quasi-degenerate states of different
angular momentum $L$ coexist.  This
produces peaks in the vortex angular momentum
dependence on $\Omega$.
We predict these peaks to broaden as $N$ increases, due to the
appearance of ``microplateaus'' which in turn lead to finite ranges
of $\Omega$ values where quasi-degenerate states coexist.

\medskip

We thank J. Dalibard, L. Pitaevskii, G.V. Shlyapnikov, and S.
Stringari for fruitful discussions.
We acknowledge support from the Deutsche
Forschungsgemeinschaft
(SFB 407, SPP 1116, GK 282, 436 POL), the EU Programme QUPRODIS, ESF
PESC QUDEDIS, EU IP Programme ``SCALA'', and the MEC (Spanish
Goverment) in terms of the contracts FIS2005-04627 and FIS2004-05639
and 2005SGR00343 from Generalitat de Catalunya.

\eject

%%%%%%%%%%%%%%%%%%%%%%%%%%%%%%%%%%%%%%%%%%%%%%%%%%%%%%%%%%%%%%%%

\begin{figure}[htb]
\includegraphics*[width=0.6\columnwidth]{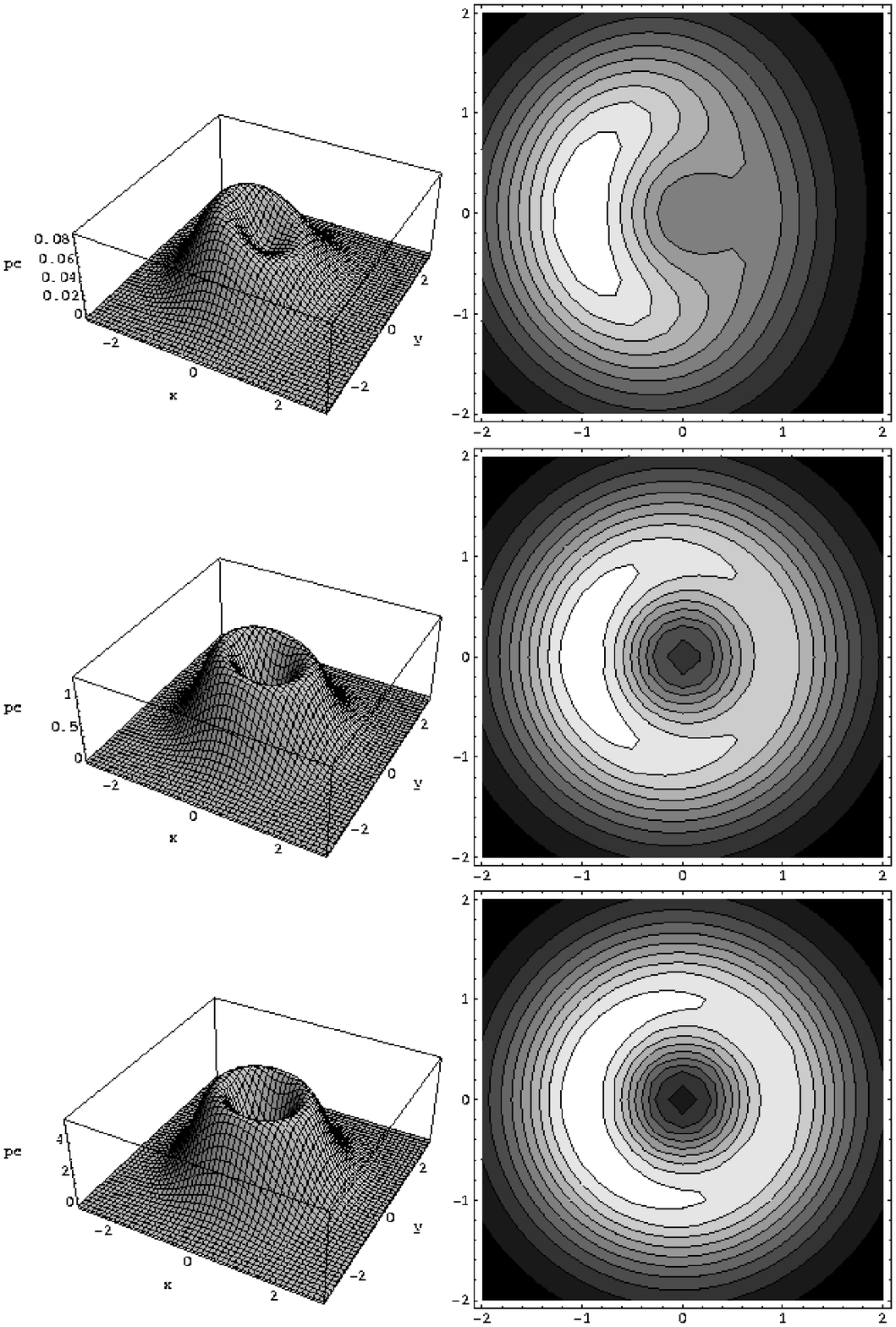}
\caption{ Pair correlation function for $N=3$, $10$ and $20$ of the
$L=N$
state. The parameter $r0$ is equal to  $0.8$, $0.95$ and $1.0$
in units of
$\lambda$ respectively.}
\end{figure}
%\eject

\begin{figure}[htb]
\includegraphics*[width=0.6\columnwidth]{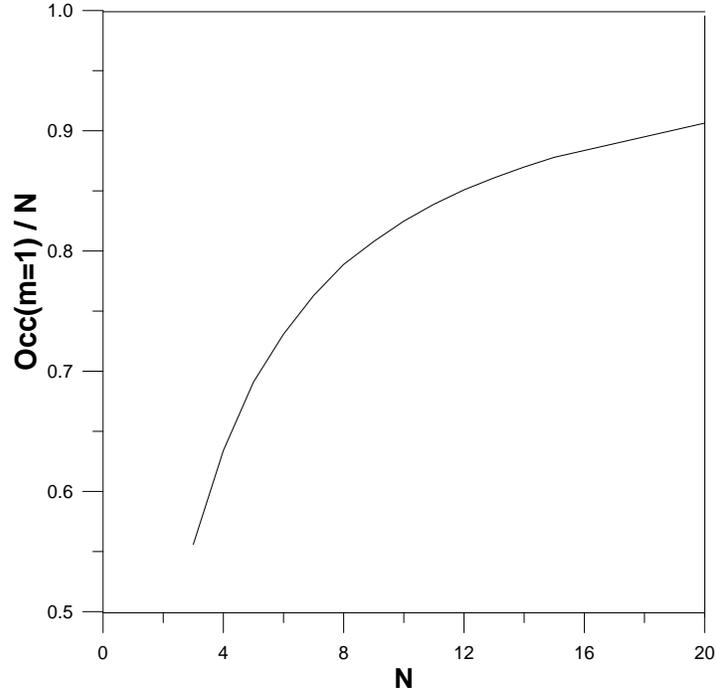}
\caption{Occupation of the m=1 single state divided by N as a function
of N.}
\end{figure}
%\eject

\begin{figure}[htb]
\includegraphics*[width=0.6\columnwidth]{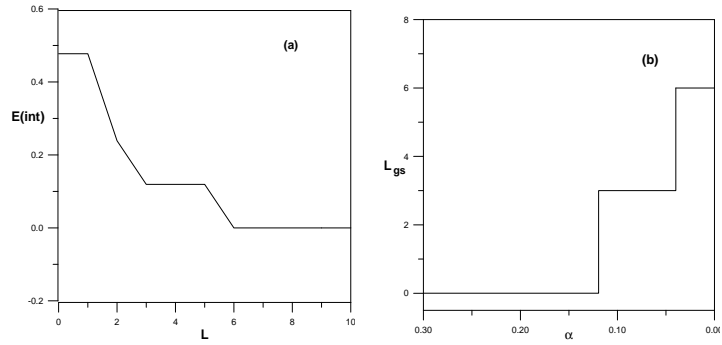}
\caption{ For $N=3$, a) interaction energy as a function of total
angular
momentum (Yrast line), b) angular momentum of the GS over $\alpha$.
The critical
values for $\alpha$ at the steps are: 0.1194 and 0.0398 in units of
$\hbar \omega_{\perp}$.}
\end{figure}
%\eject

\begin{figure}[htb]
\includegraphics*[width=0.6\columnwidth]{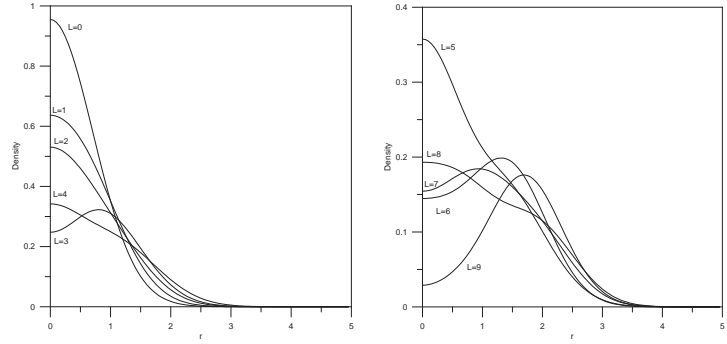}
\caption{N=3, density of the L-states}
\end{figure}
%\eject

\begin{figure}[htb]
\includegraphics*[width=0.6\columnwidth]{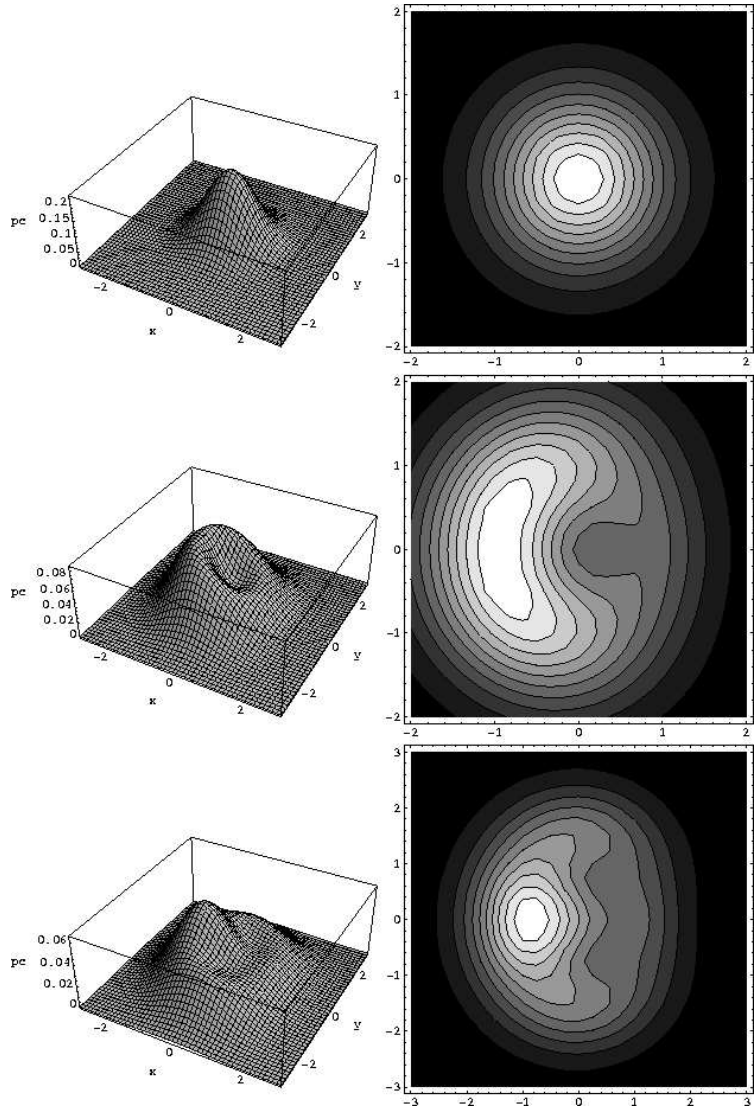}
\caption{Pair correlation function of $N=3$, for $L=0$, $3$ and $4$,
$r0=1.0$, $0.8$ and $1.0$ respectively.}
\end{figure}

\begin{figure}[htb]
\includegraphics*[width=0.6\columnwidth]{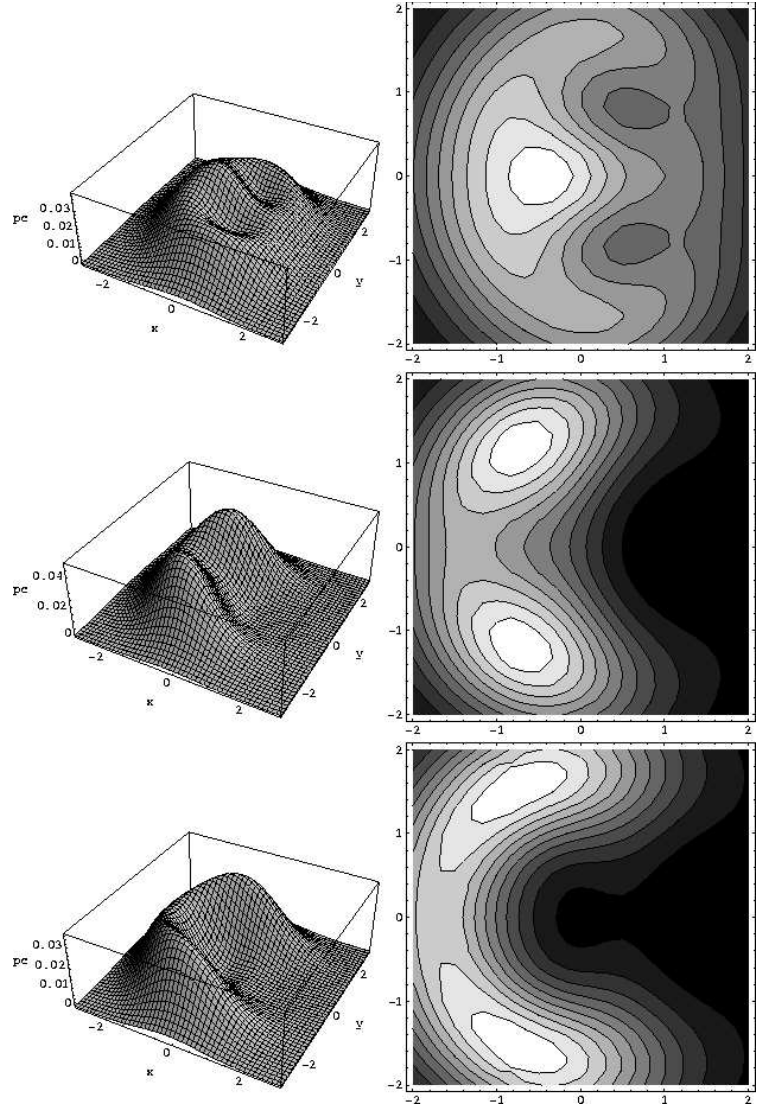}
\caption{The same as Fig.7 for $L=5$, $6$ and $9$, $r0 = 1.0$, $1.3$
and
$1.7$ respectively.}
\end{figure}

\begin{figure}[htb]
\includegraphics*[width=0.6\columnwidth]{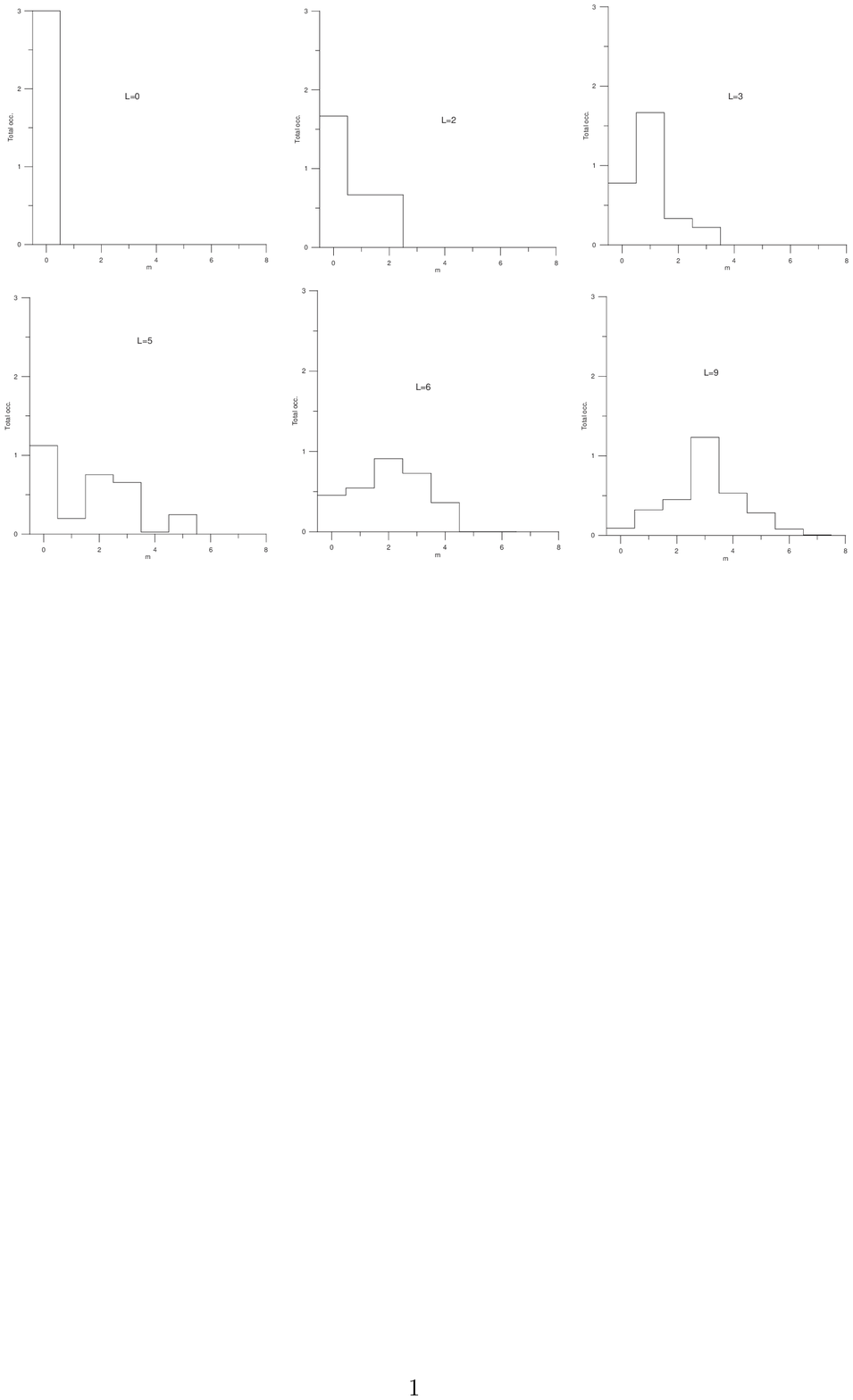}
\caption{For $N=3$, total  occupations of the single
particle states of angular momentum $m$ for several $L$-states.}
\end{figure}

\begin{figure}[htb]
\includegraphics*[width=0.6\columnwidth]{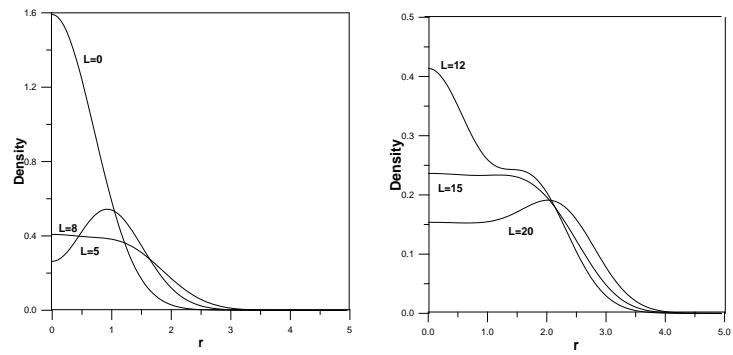}
\caption{N=5 density of the L-states (GS).}
\end{figure}
%\eject

\begin{figure}[htb]
\includegraphics*[width=0.6\columnwidth]{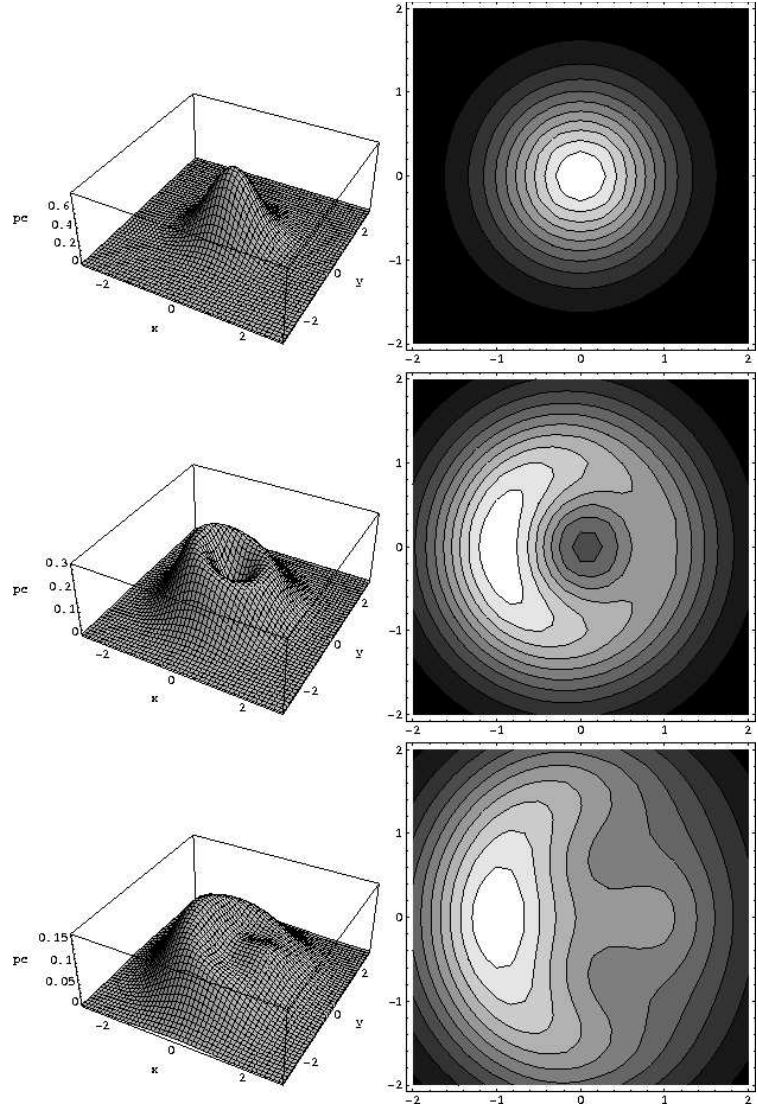}
\caption{Pair correlation of the L-states (GS) for $N=5$ and $L=0$,
$5$, and
$8$.  $r0=1.0$, $0.9$ and $1.0$
respectively.}
\end{figure}
%\eject

\begin{figure}[htb]
\includegraphics*[width=0.6\columnwidth]{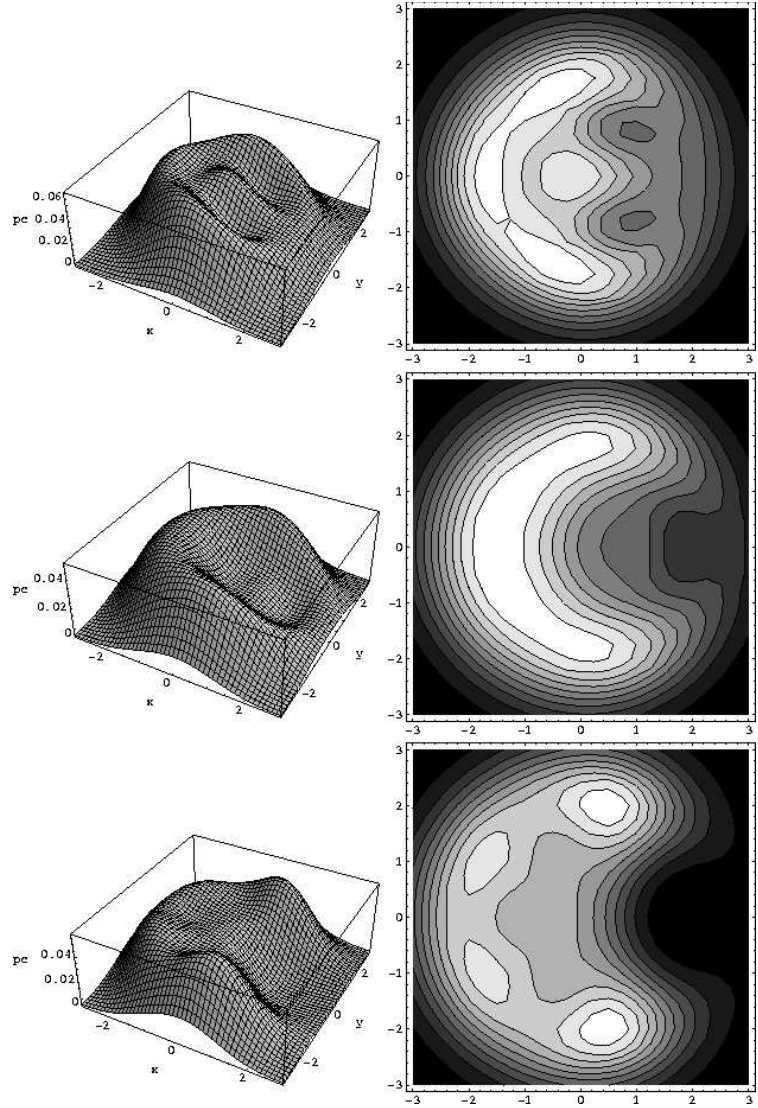}
\caption{Pir correlation of the L-states (GS) for $N=5$ and $L=12$,
$15$ and $20$.  $r0=1.0$, $1.0$ and $2.0$
respectively.}
\end{figure}
%\eject

\begin{figure}[htb]
\includegraphics*[width=0.6\columnwidth]{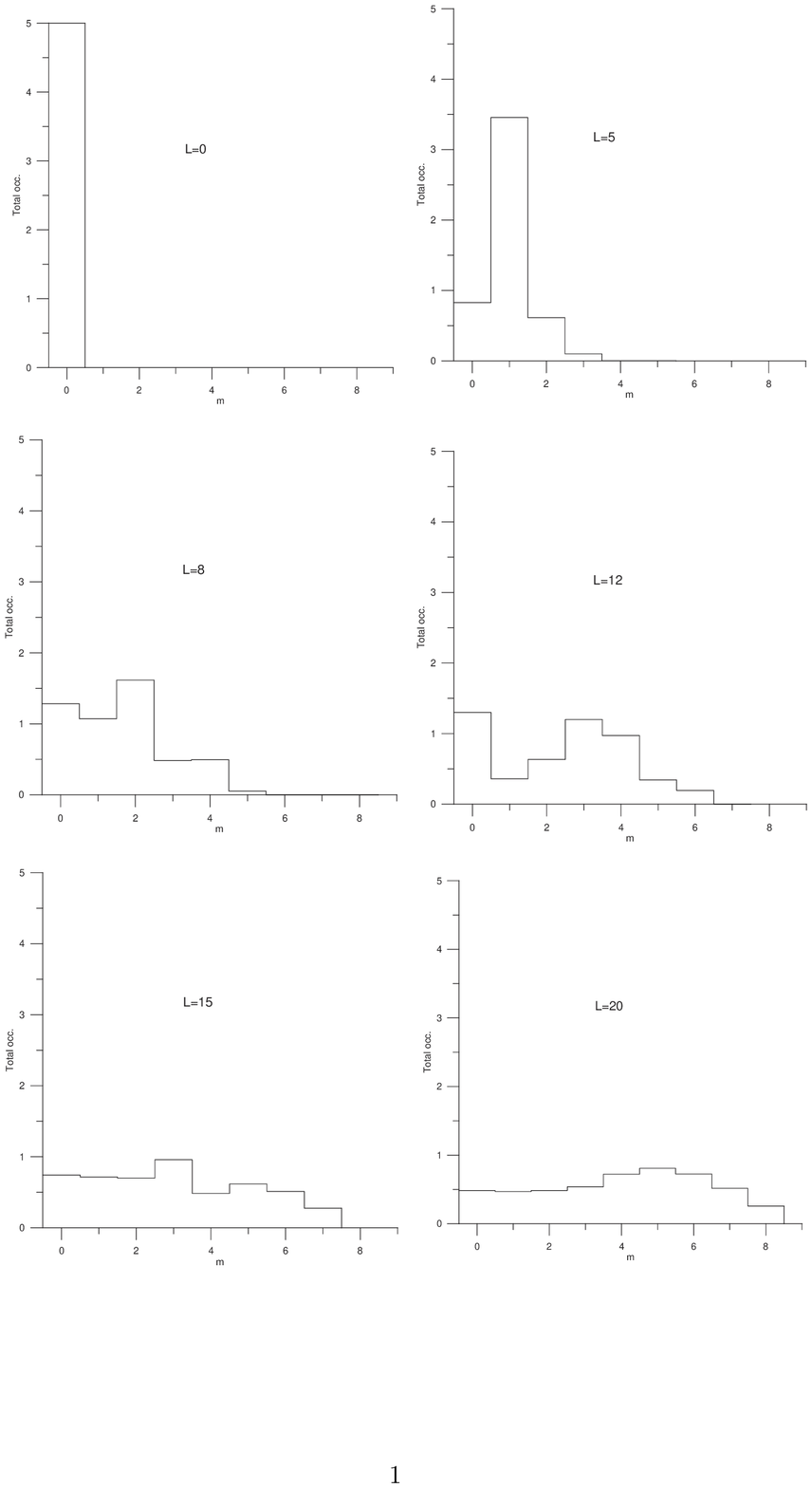}
\caption{N=5 occupations of the L-states (GS).}
\end{figure}
%\eject

\begin{figure}[htb]
\includegraphics*[width=0.6\columnwidth]{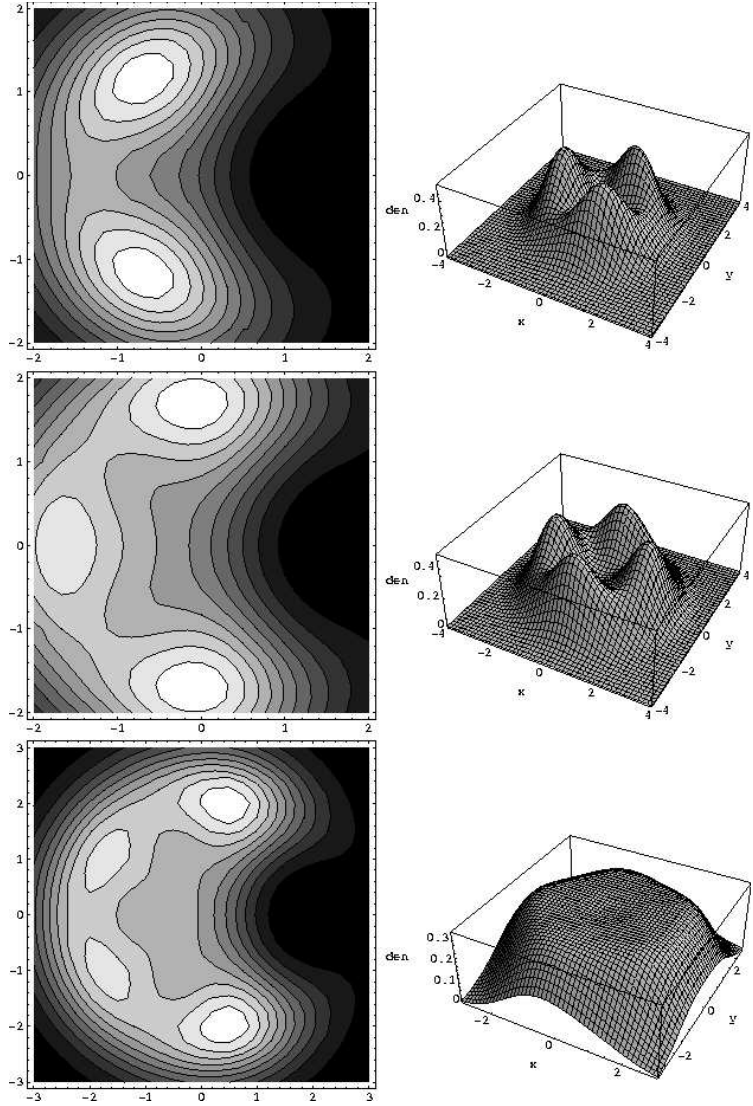}
\caption{For $N=3$, $4$ and $5$ the left hand plots show the pair
correlations (contour plots) for
$L=6$, $12$ and $20$ respectively (the Laughlin states),
with $r0=
\sqrt{N-1}$. The right hand 3D-plots show the density of the
mixtures (with equal coefficients) of angular momenta $6+9$, $12+16$
and
$20+25$ respectively.}
\end{figure}
%\eject

\begin{figure}[htb]
\includegraphics*[width=0.6\columnwidth]{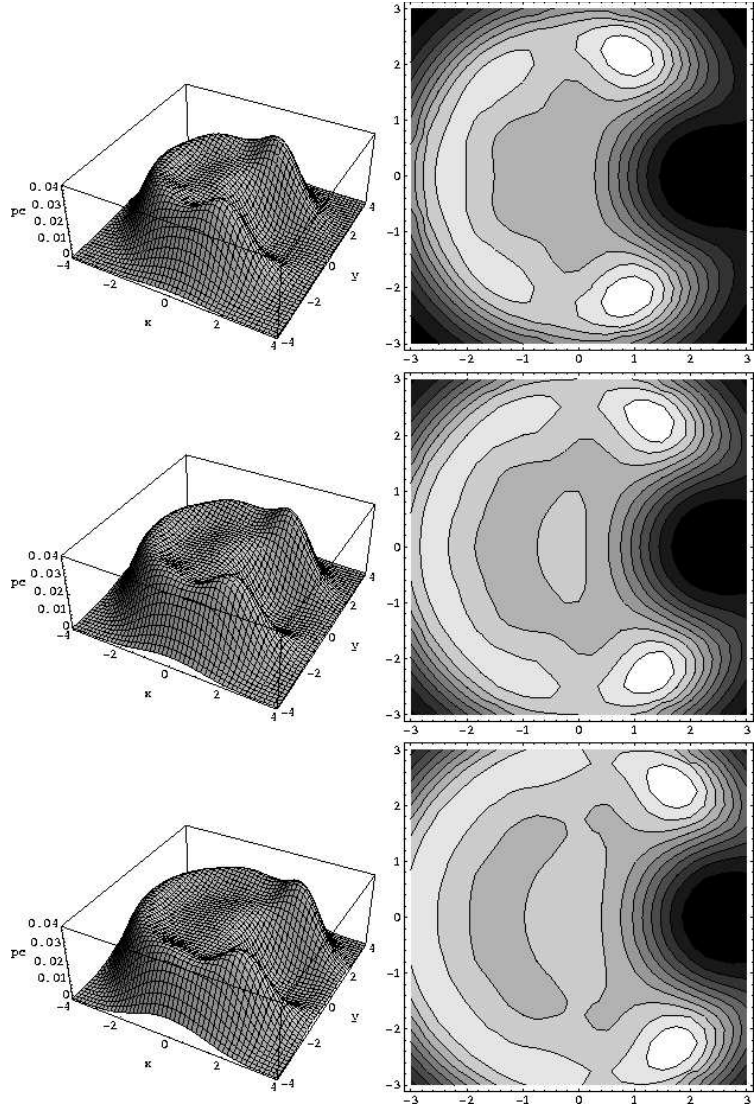}
\caption{For $N=6$, $7$ and $8$ pair correlations ( 3D-plot and
contour-plot) of
$L=30$,
$42$ and
$56$ respectively (the Laughlin states). We consider  $r0 =
\sqrt{N}$.}
\end{figure}
%\eject

\begin{figure}[htb]
\includegraphics*[width=0.6\columnwidth]{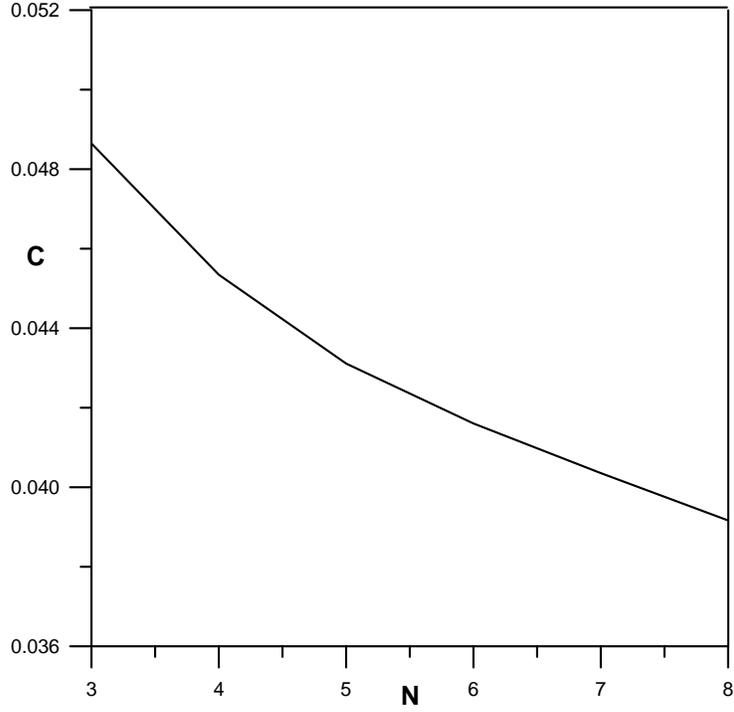}
\caption{Correlation degree (see text) of the Laughlin state as a
function of N.}
\end{figure}
%\eject

\begin{figure}[htb]
\includegraphics*[width=0.6\columnwidth]{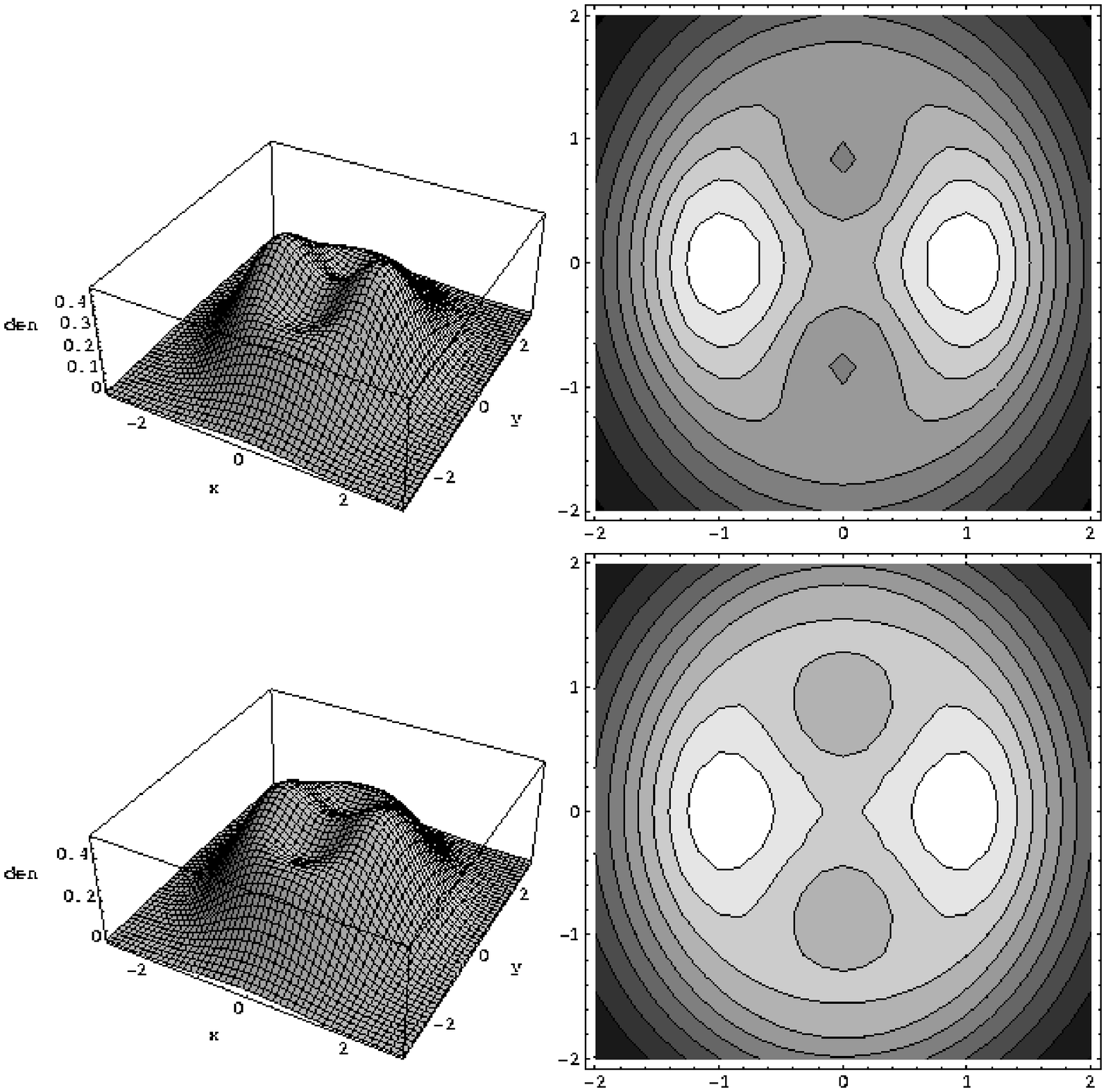}
\caption{For $N=5$ and $6$ density ( 3D-plot and contour-plot) of
the
two vortex structures. For $N=5$,  $L=8+10+12$ and for $N=6$,
$L=10+12+14$.}
\end{figure}
%\eject

%%%%%%%%%%%%%%%%%%%%%%%%%%%%%%%%%%%%%%%%%%%%%%%%%%%%%%%%%%%%%%%%%%%%%%%%%


\begin{references}
\bibitem{footnoteone}
In experiments of the ENS group, for instance,
a laser beam forms a Gaussian optical potential. It rapidly oscillates
through the trap centre along the direction that slowly rotates
around the trap centre. Obviously, such a process couples states with
various angular momenta.
\bibitem{footnotetwo}
Such a model describes the experiments of Ref.~\cite{mad}
quite well, and has several different advantages:
i) it has a simple analytical form and allows for simple calculations;
ii) it also describes other kinds of ``stirring'', for instance,
the one in which one uses an elliptical, rotating mask to
generate the optical potential of Eq.~(\ref{aniso}).
\end{references}
\end{document}